\documentclass[fleqn]{aa}
\usepackage{graphicx, natbib,epsfig,amsmath,amssymb, mathrsfs, txfonts}
\DeclareGraphicsExtensions{.eps, .jpg, .ps}
\usepackage{epstopdf}
\bibpunct{(}{)}{;}{a}{}{,}
\begin{document}

\title{Apodized-pupil Lyot coronagraphs: \\multistage designs for extremely large telescopes}
\author{P. Martinez\inst{1}}
\institute{European Southern Observatory, Karl-Schwarzschild-Strasse 2, D-85748, Garching, Germany}
\offprints{P. Martinez}

\abstract
{Earlier apodized-pupil Lyot coronagraphs (APLC) have been studied and developed to enable high-contrast imaging for exoplanet detection and characterization with present-day ground-based telescopes. 
With the current interest in the development of the next generation of telescopes, the future extremely large telescopes (ELTs), alternative APLC designs involving multistage configuration appear attractive. }
{I study the relevance of these designs for the ELTs. My analysis is designed to find out the implications of inherent error sources occurring on large coronagraphic telescopes to the performance of multistage APLC configurations.}
{Performance and sensitivity of multistage APLC to ELT specificities are analyzed and discussed, taking into account several ineluctable coronagraphic telescope error sources by means of numerical simulations.  
Additionally, a first laboratory experiment with a two-stages-APLC in the near-infrared ($H$-band) is presented to further support the numerical treatment.} 
{Multistage configurations are found to be inappropriate to ELTs. The theoretical gain offered by a multistage design over the classical single-stage APLC is largely compromised by the presence of inherent error sources occurring in a coronagraphic telescope, and in particular in ELTs. The APLC remains an attractive solution for ELTs, but rather in its conventional single-stage configuration. }
{}


\maketitle

\section{Introduction}
Direct imaging and spectroscopy of extrasolar planets is an exciting and ambitious goal, which will considerably expand with the imminent rise of powerful instruments at the VLT \citep[SPHERE,][]{SPHERE}, Gemini \citep[GPI,][]{GPI}, and Subaru \citep[HiCIAO,][]{HiCIAO} observatories.
These ground-based adaptive optic instruments built for 8 m class telescopes include advanced starlight cancellation techniques such as coronagraphy to deliver contrast ratios of about $10^{-6}$ to $10^{-7}$ at a few arcsecond in the near-IR.
Typically, detection and spectroscopic characterization of relatively young objects, giant planets, and brown dwarfs, will be possible around nearby stars. 
With the worldwide emergence of ELT projects older and fainter companions out of the sight of current and imminent instruments will become accessible. These future extremely large telescopes will improve the sensitivity of exoplanet searches toward lower masses and closer angular distances, ideally down to rocky planets.

Extremely large telescopes will provide an increase of resolution and contrast that will enable us to explore areas that at present 8 to 10 m telescopes cannot resolve. Coronagraph designs and optimization must be addressed accordingly.
With these considerations in mind, we have previously studied a trade-off for coronagraphy in the context of ELTs \citep{Corono}, to initiate a general comparison of coronagraphs, with the goal to identify valuable concepts and fields of applications. Among them the apodized-pupil Lyot coronagraphs \citep[APLC, ][]{2002A&A...389..334A, 2003A&A...397.1161S} was identified as an adequate baseline solution.
\begin{figure*}[!t]
\begin{center}
\includegraphics[width=16cm]{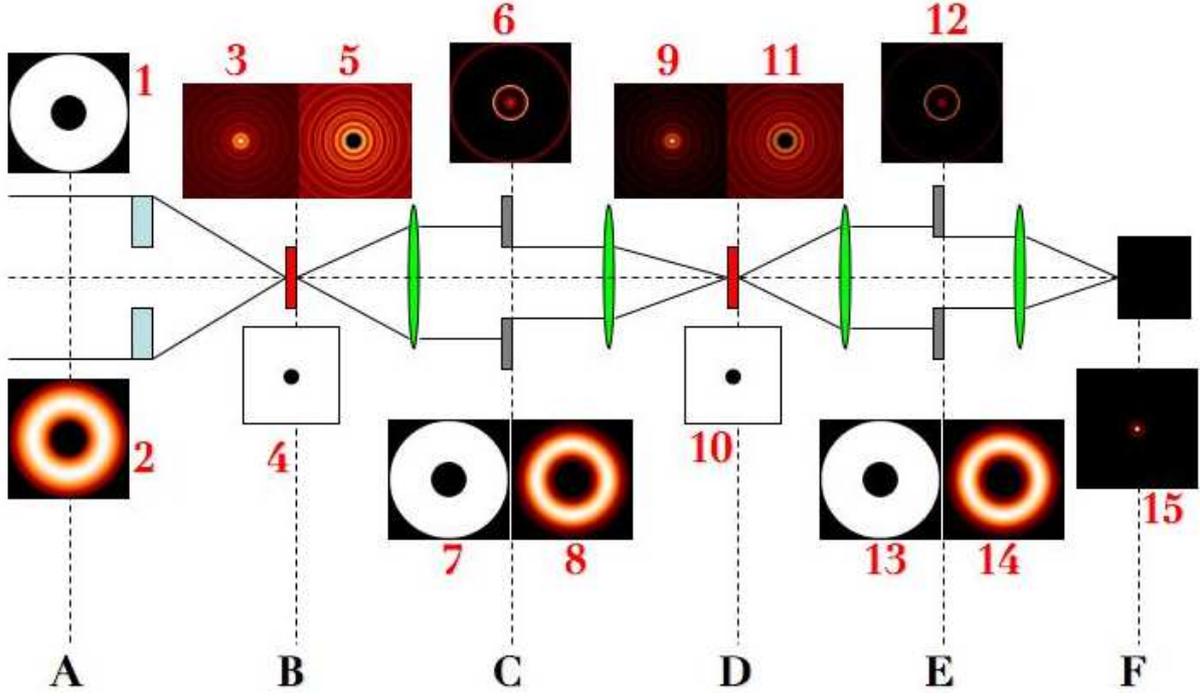}
\end{center}
\caption{MAPLC ($n$=2) coronagraphic process illustration from the telescope pupil plane (A) to the detector plane (F).}
\label{fig1}
\end{figure*} 
The APLC is one of the most advanced concepts studied intensively in the past few years \citep{2002A&A...389..334A, 2003A&A...397.1161S, 2005ApJ...618L.161S, 2007A&A...474..671M, Corono, 2009ApJ...695..695S}, which has been successfully developed and tested in the laboratory \citep{2008SPIE.7015E.175M, APLCtest, Reynard, microdots, microdots2}. It will very likely soon provide scientific results on the sky through several projects (e.g. SPHERE, GPI). In \citet{2004SPIE.5490..456A}, and more recently in the context of ELTs \citep{2009ApJ...695..695S}, more sophisticated APLC implementations have been proposed. 
Among them the interesting possibility of multistage APLC has been presented with the goal of providing deeper contrast, and/or to relax the inner working angle (IWA).

As discussed in \citet{2009ApJ...695..695S}, the features of ELT geometry, in particular the large secondary support structures, divide the APLC parameter space trade-off for ELTs into two philosophies:
(1/) to select small mask-size configurations ($\lesssim$ 5$\lambda/D$ diameter) to neglect the secondary support presence, and thereby use a rotationally symmetric apodizer, (2/) implement larger mask sizes ($\gtrsim$ 5$\lambda/D$) combined with more complicated apodizers. The latter provides better contrast but a larger IWA, and involves complex structures in the apodizer with uncertainties related to both manufacturing and practical aspects (e.g. alignment, pupil rotation). To preserve the advantage of the natural increase of resolution provided by the ELTs, smaller mask sizes are preferable. In this context multistage APLC designs (MAPLC hereafter) could gather all adequate advantages either by delivering improved contrast, or by enabling smaller IWA, through avoiding complex structures in the apodizer. However, the advantages of MAPLC over the standard APLC have not been explored in a systematic way. 

This study is therefore largely motivated by the need to determine whether  MAPLC is a suitable solution for ELTs, and to ascertain the robustness of the gain offered by this design.
The paper proposes a detailed study of MAPLC by considering ELTs specificities and inherent error sources occurring in a coronagraphic telescope. A comparison with a conventional single-stage APLC is carried out.
In Sect. 2 the formalism of APLC is recalled and an extension to the multistage case is provided. Section 3 provides the simulation analysis and results, while in Sect. 4 a first laboratory demonstration in the near-IR of a two-stages APLC is presented to further confirm the numerical treatment. Finally in conclusions are drawn Sect. 5.

\section{Multistage apodized pupil Lyot coronagraph}

\subsection{Formalism}
In this section, the formalism of APLC is recalled following the notations of \citet{2007A&A...474..671M}. From this basis, a generalization for multistage configurations is provided. 
The APLC is by principle a combination of a classical Lyot coronagraph (hard-edged occulting focal plane mask, FPM hereafter) with an apodization of the entrance aperture.
The MAPLC ($n$ stands for the number of stages) has the advantage that the field of the pupil stop of a single-stage APLC ($n$=1) is naturally apodized, and can  therefore be used as an entrance pupil for a second stage, and so on. Multistage APLC $n$ stages uses $n$ identical FPMs and pupil stops in a cascading mode, with only one unique pupil apodizer. Consequently no further loss in transmission occurs to that of a single-stage APLC.   

Below, the variables $r$ and $\rho$ (for pupil plane and focal plane respectively) are omitted for the sake of clarity.   
The function that describes the FPM is noted $M$ (equal to 1 inside the coronagraphic mask and to 0 outside). The FPM is then equal to $1 - M$.
While $P$ is defined as the telescope aperture function, $\phi$ stands for the apodizer function. $\Pi$ describes the pupil stop function, which is considered in a first approximation to be equal to the telescope aperture ($\Pi = P$, in principle the APLC does not require further pupil-stop reduction, but a slight reduction is mostly carried out for alignment issues).
The Fourier transform of a function $f$ is noted $\hat{f}$, while the symbol $\otimes$ denotes the convolution product. 

The study is restricted to the case of $n$=2 (see Fig. \ref{fig1}), sufficient to provide a generalization to MAPLC. It is assumed that pupil-stops and FPMs remain identical in all $n$ planes (\citet{2009ApJ...695..695S} provides a mathematical justification of this condition).
Planes A, B, C, D, E and F (Fig. \ref{fig1}, and Eq. \ref{APLC2-1} to Eq. \ref{amplitudecorono2}) correspond to the telescope aperture, the coronagraphic first focal plane, the pupil-stop plane, the second coronagraphic focal plane, the second pupil plane, and the detector plane respectively. Below, numbers in [] refer to their corresponding images of Fig. \ref{fig1}.

\noindent The entrance pupil is apodized in the pupil plane [1, 2]
\begin{equation}
\psi_A =  P\phi.
\label{APLC2-1}
\end{equation}
\noindent The complex amplitude of the star [3] is spatially filtered ([5], for the low frequencies) by the first FPM [4]
\begin{equation}
\psi_B = \hat{\psi}_A \times [1 -  M].
\end{equation}
\noindent The exit pupil image [6] is spatially filtered (high frequencies) by the first pupil-stop [7]
\begin{equation}
\psi_C = \hat{\psi}_B \times \Pi,
\end{equation}
\begin{equation}
\psi_C = [\psi_A -  \psi_A \otimes \hat{M}] \times \Pi,
\label{EqC}
\end{equation}
\noindent and as a result the relayed pupil is attenuated [8] and proportional to the apodized entrance aperture.
\noindent The coronagraphic complex amplitude [9] before the next focal plane mask (denoted $\psi_{D^{-}}$) is  
\begin{equation}
\psi_{D^{-}}  = \hat{\psi}_C = [\hat{\psi}_A  -  \hat{\psi}_A M]  \otimes \hat{\Pi}. 
\label{firststage}
\end{equation}
\noindent The coronagraphic complex amplitude [9] is not imaged on the detector but is once more spatially filtered [11] by a second FPM [10] (for the sake of clarity, and to avoid confusion with Eq. \ref{firststage}, $\psi_{D}$ is here denoted $\psi_{D^{+}}$ to reflect to action of the second FPM):
\begin{equation}
\psi_{D^{+}}  = \left( [\hat{\psi}_A  -  \hat{\psi}_A M]  \otimes \hat{\Pi} \right) \times [1 -  M].
\end{equation}
 \noindent The second exit-pupil image [12] is spatially filtered for the high frequencies [14] by a second pupil-stop [13]
\begin{equation}
\psi_E = \hat{\psi}_{D^{+}} \times \Pi,
\label{EqE}
\end{equation}
\noindent where $\hat{\psi}_{D^{+}}$ is expressed as
\begin{equation}
\hat{\psi}_{D^{+}} = \left( [\psi_A -  \psi_A \otimes \hat{M}] \times \Pi \right) -  \left( [\psi_A -  \psi_A \otimes \hat{M}] \times \Pi \right) \otimes \hat{M}, 
\label{EqD}
\end{equation}
\noindent and using Eq. \ref{EqC} into Eq. \ref{EqD}, one can simplify Eq. \ref{EqE} to
\begin{equation}
\psi_E = (\psi_C -  \psi_C \otimes \hat{M}) \times \Pi.
\end{equation}
\noindent The coronagraphic amplitude on the detector plane [15] becomes
\begin{equation}
\psi_F = \hat{\psi}_E = (\hat{\psi}_C -  \hat{\psi}_C  \times M) \otimes \hat{\Pi},
\end{equation}
\noindent which finally can be expressed as
\begin{equation}
\psi_F  = (\psi_{D^{-}} -  \psi_{D^{-}}  \times M) \otimes \hat{\Pi}.
\label{amplitudecorono2}
\end{equation}
\noindent From Eq. \ref{amplitudecorono2} it is straightforward to see that each stage produces a similar attenuation of the starlight (assuming that FPMs are identical in the $n$ stages).
Therefore, one can generalize the result as
\begin{equation}
\psi_{n}  = (\psi_{n-1} -  \psi_{n-1}  \times M) \otimes \hat{\Pi},
\label{final}
\end{equation}
\noindent where $\psi_{n}$ corresponds to the final coronagraphic amplitude in the detector plane of a $n$ stages APLC design.
Following the notation of \citet{2004SPIE.5490..456A}, the coronagraphic efficiency (total energy attenuation) of single-stage APLC being $Q$, the efficiency of $n$ stages APLC is in principle $Q^{n}$.

\subsection{Advantages of MAPLC}
Multistage APLC arise as attractive solutions because by principle (i.e. in ideal conditions) each successive coronagraphic stage provides the same rejection factor of the starlight, and the whole design only involves one physical apodizer-component at the entrance of the system.  As a result, better contrasts are available with a multistage design for a given APLC configuration, while the system transmission is preserved through the successive stages (i.e. no further reduction of the throughput by the successive stages). 
On the other hand, a MAPLC can be designed to improve the IWA of a single stage APLC by maintaining high/similar performance.
As APLCs can be designed for any arbitrary telescope aperture \citep{2005ApJ...618L.161S}, MAPLCs are applicable whatever the central obscuration ratio of the telescope.


A first straightforward limitation of MAPLC designs comes from the chromaticity, unless achromatic APLC solutions can be obtained \citep{2005PASP..117.1012A}.
The efficiency of a MAPLC relies on the proportionality property between the entrance apodized-pupil and the successive pupil planes. The $n$ pupil planes require the field amplitude to be proportional the apodizer function $\phi$. This condition is only satisfied at a single wavelength, not in broadband conditions. For instance, the performance degradation (total energy rejection) to that of a monochromatic case (denoted $\mathscr{D}$) is of about 10$\%$ for a single-stage APLC when $\lambda \slash \Delta \lambda = 5\%$. In similar conditions, a two-stages APLC degrades even more: $\mathscr{D}= 94\%$. Achromaticity of the APLC is mandatory to take advantage of the multiple stages implementation.

\section{Sensitivity analysis}
In this section the impact of several parameters on a MAPLC ($n$=2, i.e. two stages) is analyzed and a comparison to that of a single-stage APLC ($n$=1) is provided.
All simulation hypotheses are presented in Sect. \ref{Assumptions}, while results are discussed in Sect. \ref{Results}. To discern the error source influence, all parameters are simulated independently from each other, and simulations are monochromatic. In Sect. \ref{application} a performance comparison ($n$=2 vs. $n$=1) is provided for a simultaneous consideration of several error sources. 

\subsection{Assumptions}
\label{Assumptions}
Simulations make use of simple Fraunhofer propagators between pupil and image planes, which are implemented as fast Fourier-transforms (FFTs) generated with an IDL code. 
The image plane is sampled with 0.125$\lambda/D$ per pixel.
As a baseline an ELT with 30\% (linear) central obscuration ratio as expected for the European-ELT \citep[E-ELT,][]{E-ELT2008} is assumed. 
As for the wavelength, a baseline of $\lambda=1.6\mu m$ (center of the H-band) is adopted. The APLC is a 4.7$\lambda/D$ based on a former study \citep{2007A&A...474..671M}, and representative of the small mask-size domain ($<$ 5$\lambda/D$) we are concerned with here.
The impact of several important parameters (non-exhaustive list) is assessed, and a brief description of their hypotheses is discussed in the next subsections.

\begin{figure*}[!ht]
\centering
\includegraphics[width=7.9cm]{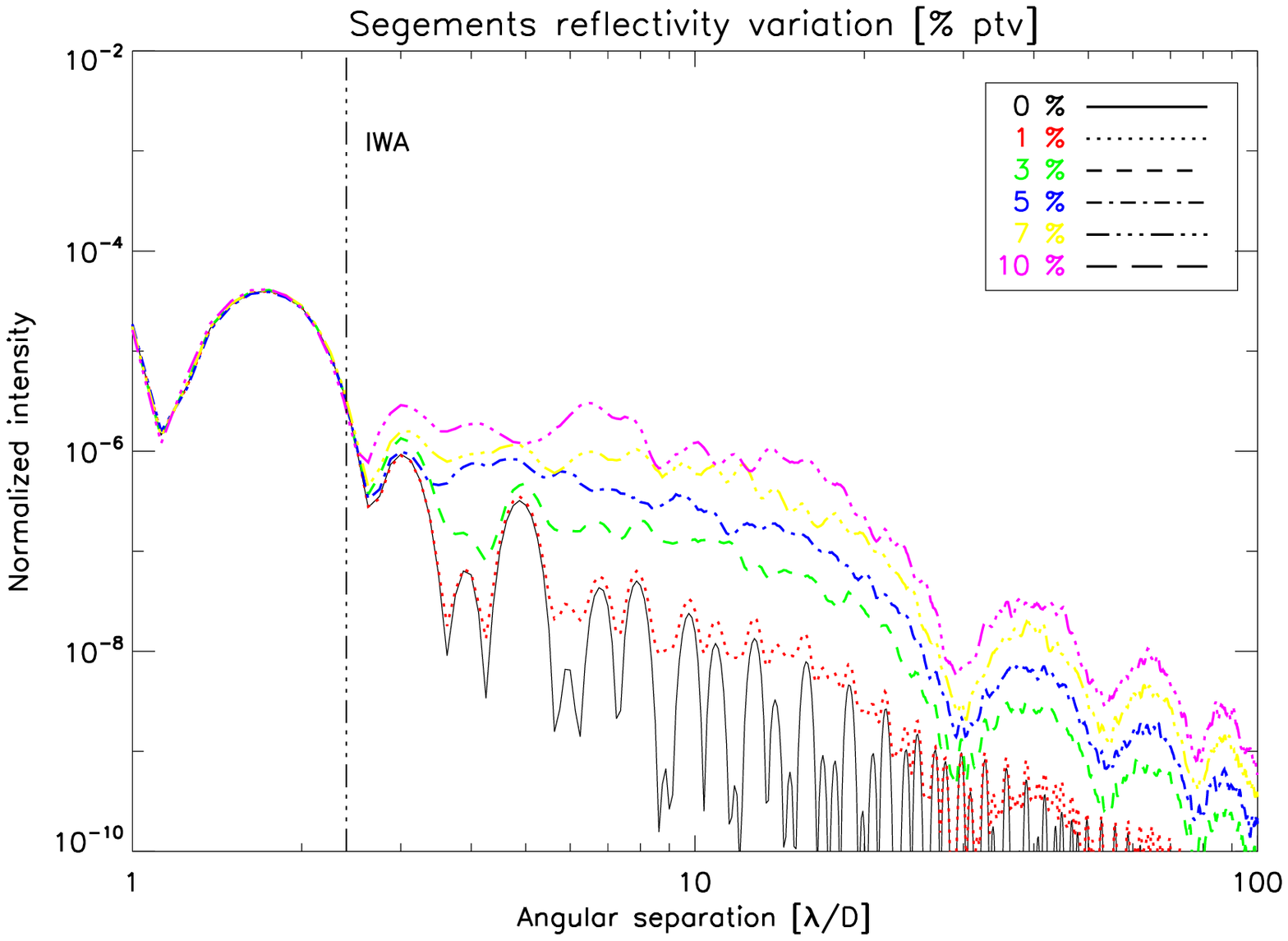}
\includegraphics[width=7.9cm]{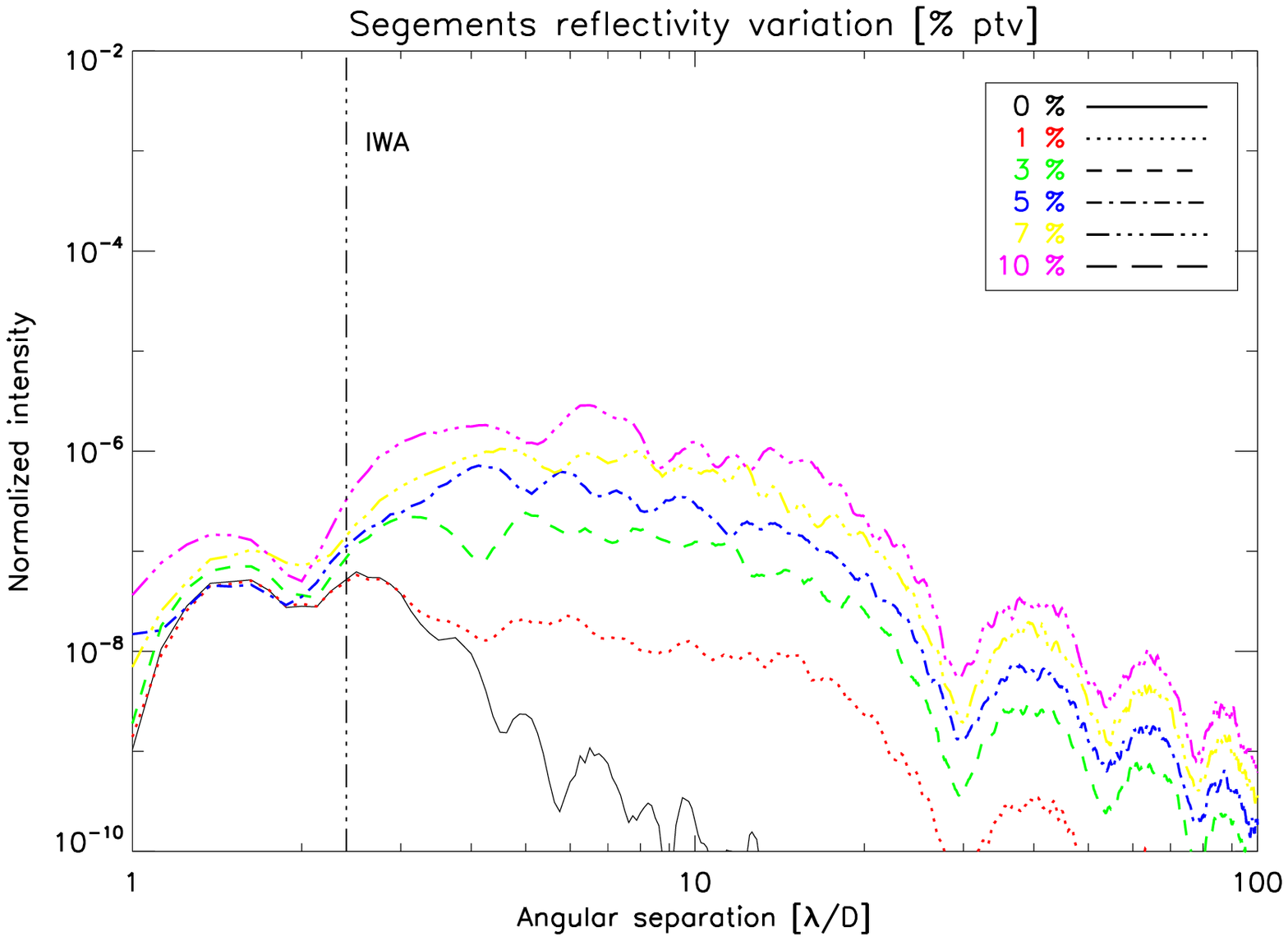}
\includegraphics[width=7.9cm]{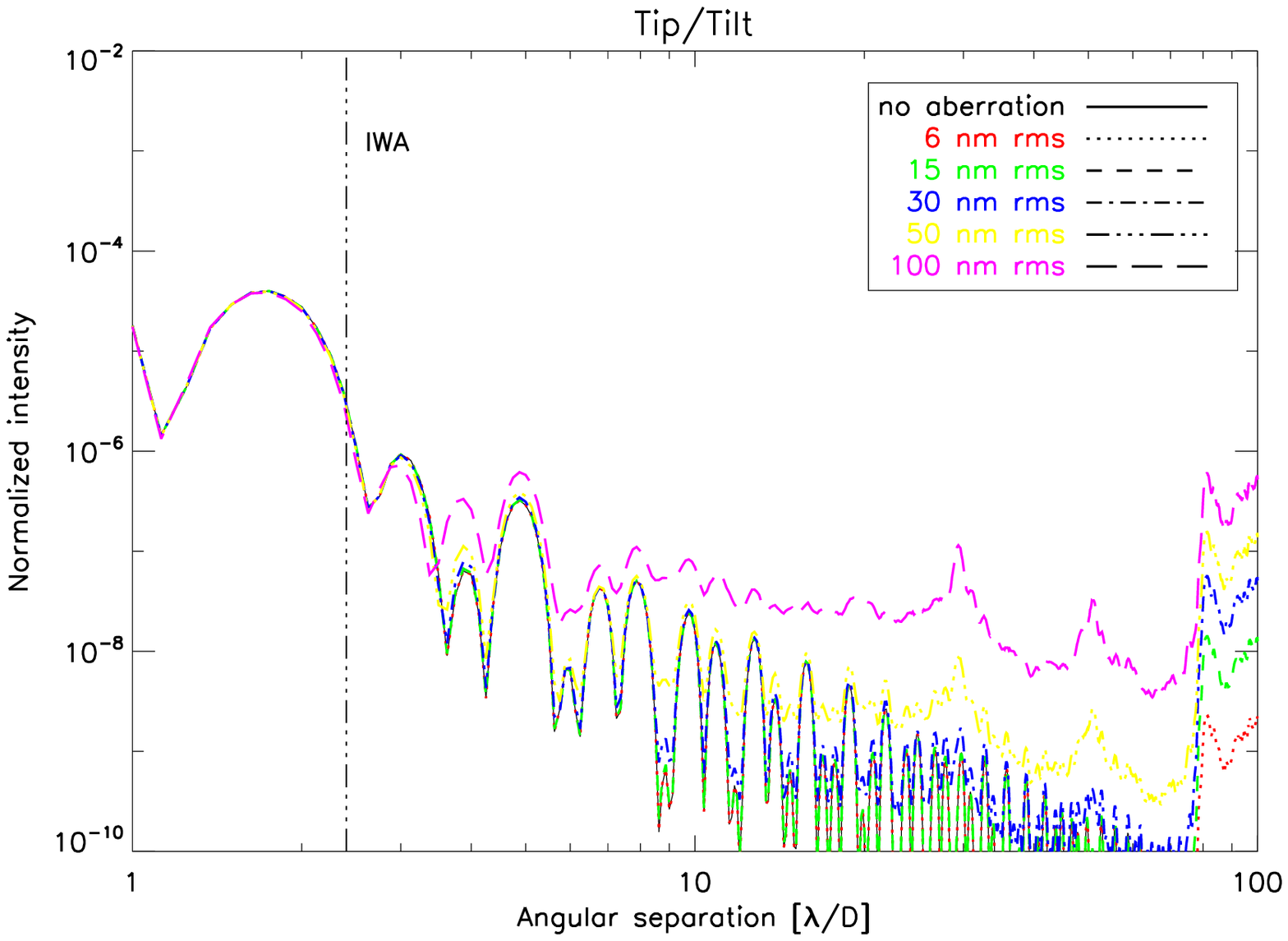}
\includegraphics[width=7.9cm]{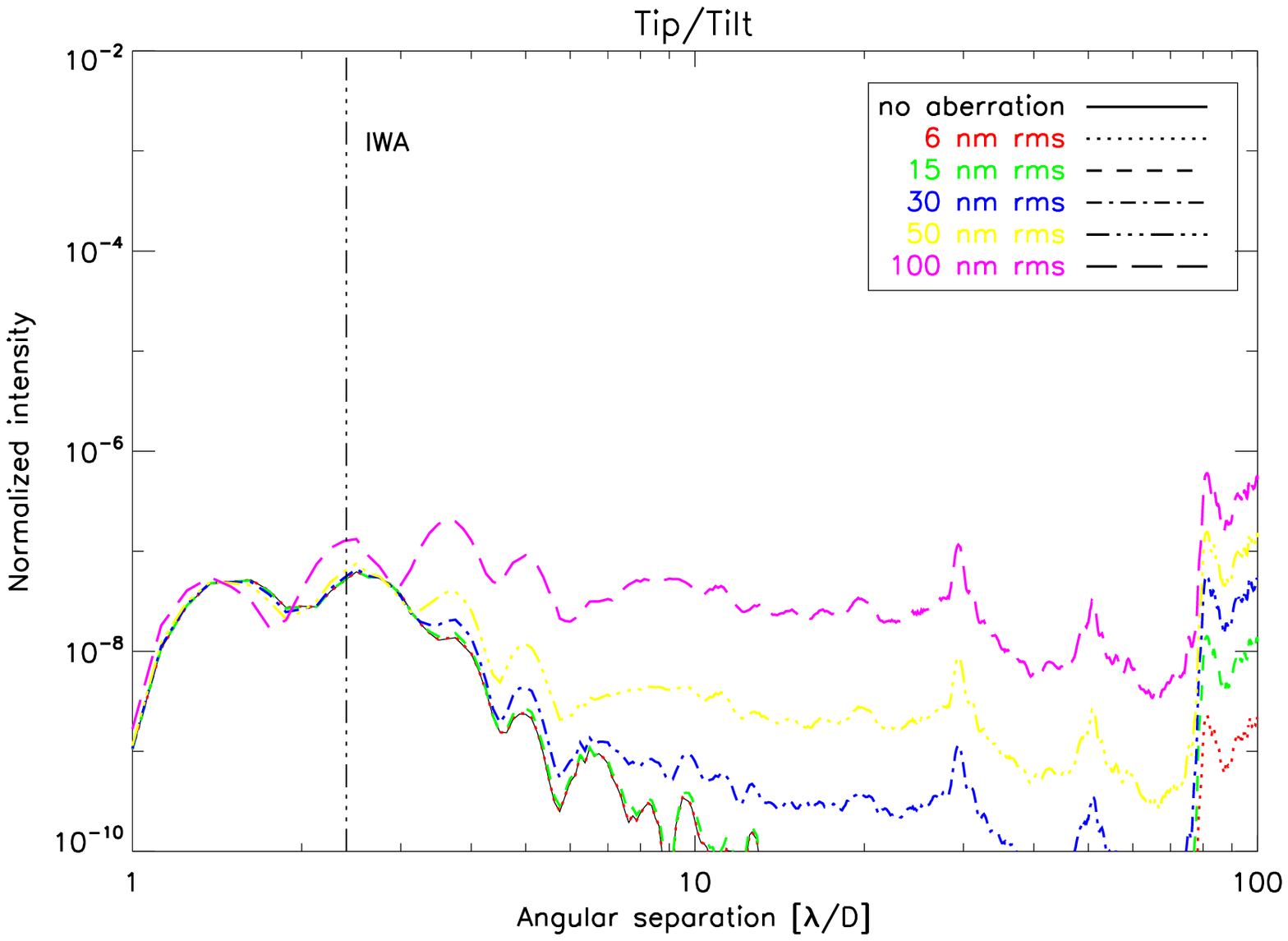}
\includegraphics[width=7.9cm]{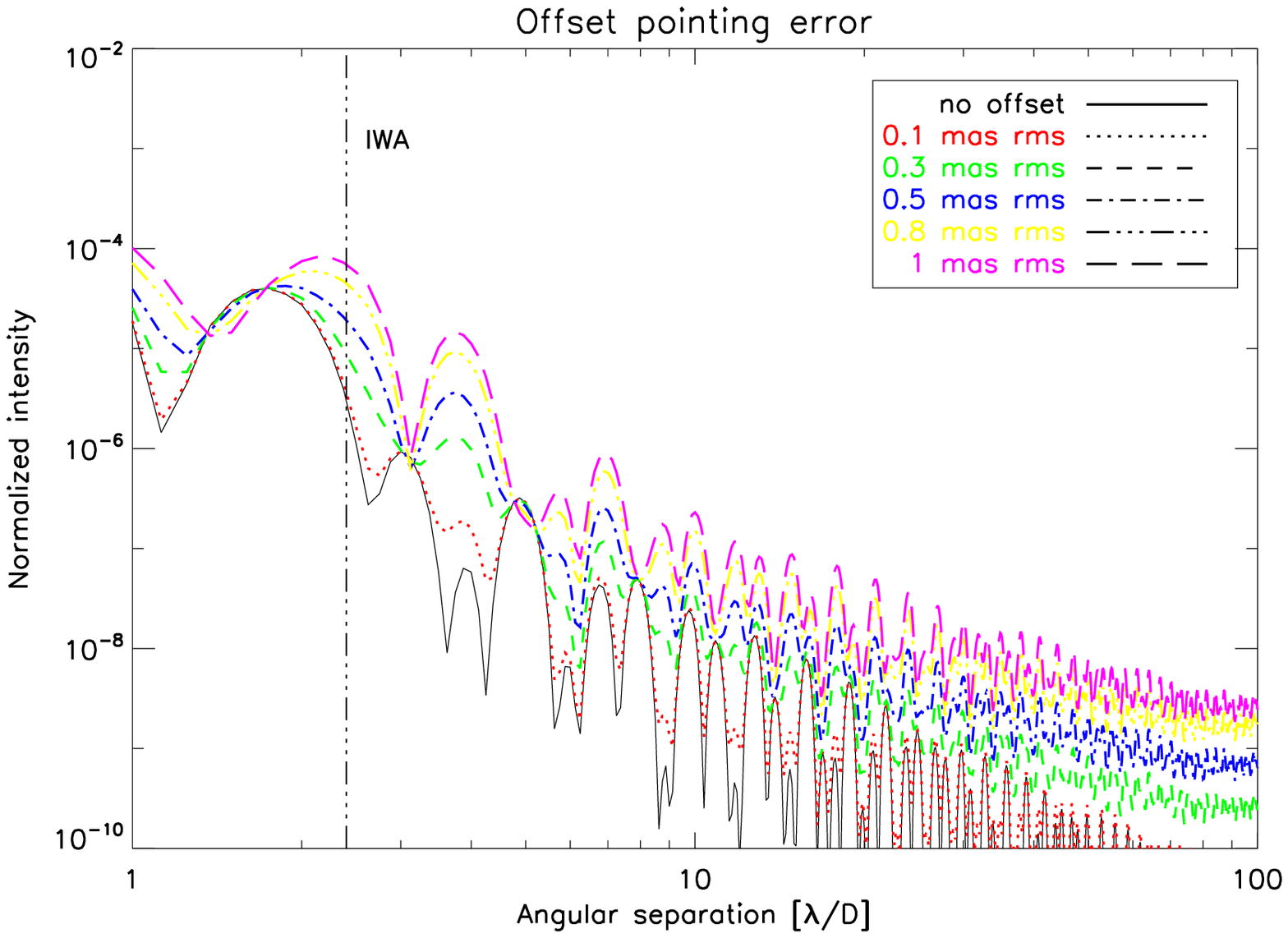}
\includegraphics[width=7.9cm]{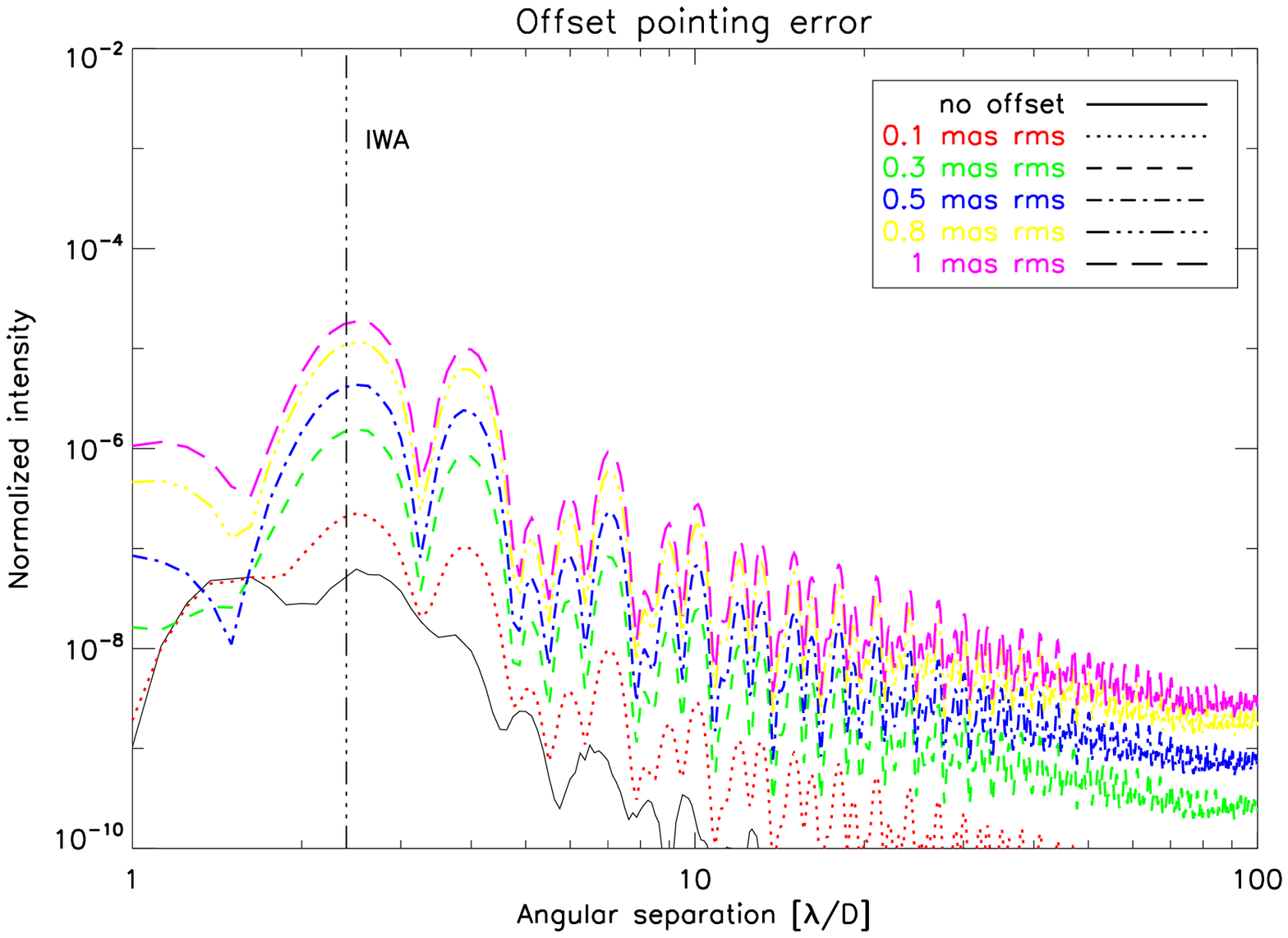}
\includegraphics[width=7.9cm]{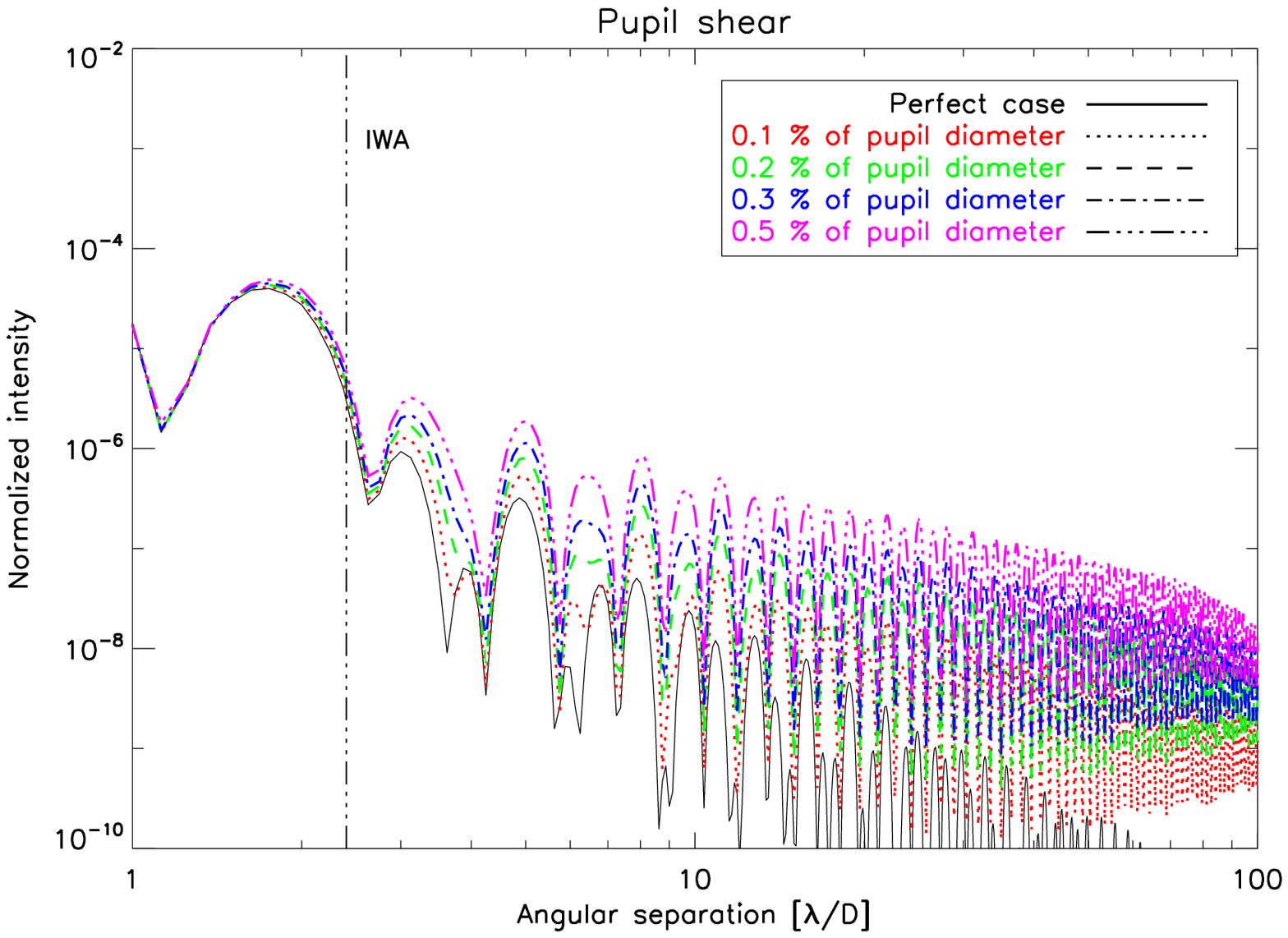}
\includegraphics[width=7.9cm]{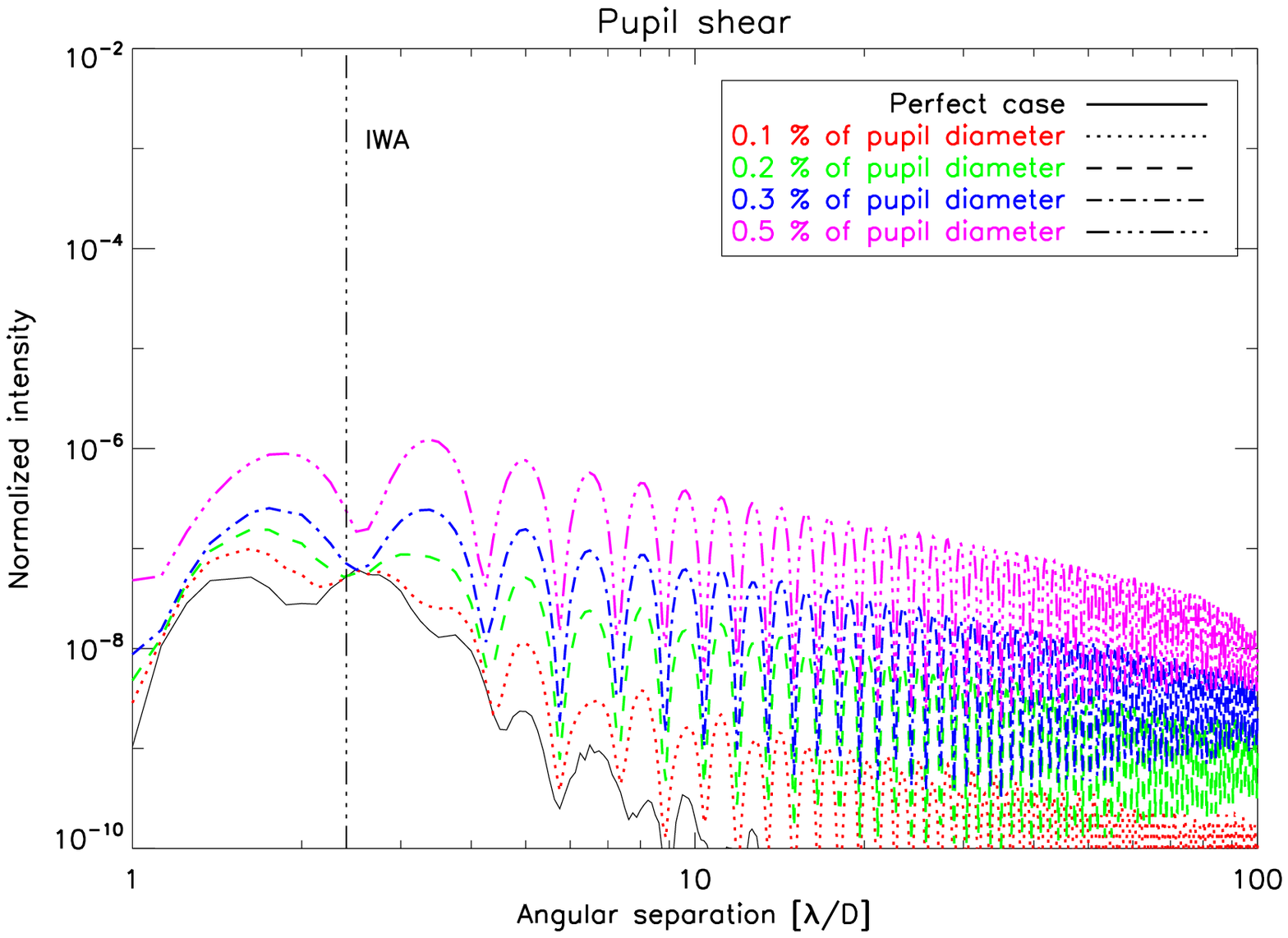}
\caption{Left column: single-stage APLC ($n$=1), Right column: two-stages APLC ($n$=2). From the top row to the bottom row: impact of the segment reflectivity, segment aberrations,  offset pointing, and pupil shear.}
\label{APLC2error}
\end{figure*}

\subsubsection{Primary mirror segmentation}
Any ELT primary mirrors will necessarily be segmented, and variations in both amplitude and phase are expected.
Amplitude variation due to a difference in reflectivity between the segments (optical coating) induces a wavefront-amplitude variation that leads to potentially bright static speckles in the focal plane of the instrument. 
Segment aberrations refer here to low-order static aberrations (piston, tip-tilt, defocus, and astigmatism), which produce speckles that fall relatively near the central core of the image. 
It is important to know how robust a coronagraph is to these defects. Here, $\sim750$ hexagonal segments of 1.5 meters diameter are assumed. 

The impact of a uniform segment-to-segment reflectivity variation from 1\% to 5\% (peak-to-valley, hereafter ptv) is assessed. For comparison, 5\% (ptv) is the typical variation measured on the Keck telescope \citep{2003SPIE.4840...81T}.

For the static aberrations, initial values are considered, while it is assumed that they are lates corrected by an XAO system (extreme adaptive optics system).
In practice the power spectral density (PSD) of the phase is set to a null contribution at frequencies lower than the XAO cut-off frequency (80$\lambda/D$, assuming 26 cm inter-actuator spacing for a 42m telescope diameter, as assumed in \citet{Corono}). This method gives the best possible correction that can be obtained (i.e. it does not include any wavefront sensor errors, influence function, nor actuators space positioning).
The XAO cut-off frequency varies depending on the XAO parameters. The simulations here are only examples and should be further adapted to each XAO system if an exact result is sought. 
A practical metric to evaluate the quality of the segment aberration correction is provided by the ratio of the corrected rms wavefront error to the initial uncorrected rms wavefront error:
\begin{equation}
\mathscr{\gamma}_{AO} = \frac{\sigma_{corrected}}{\sigma_{initial}}.
\label{AOequ}
\end{equation}
With the simple PSD shaping, the following values for $\mathscr{\gamma}_{AO}$ are obtained: 0.22, 0.34, 0.27, 0.41 for piston, tip-tilt, defocus, and astigmatism, respectively, which is close to the analytical expectation \citep{2006SPIE.6267E..86Y}.
Former measurements with the Keck telescope \citep{2000SPIE.4003..188C} show that 10 nm rms is reachable. 
A range of intial wavefront-errors from 6 to 100 nm rms is considered in the simulation, which is corrected by the XAO system, and hence reduced (tip-tilt case) to values ranging from 2 to 34 nm rms.
In practice, each static aberration is analyzed independently. Limited differences in the coronagraphic halo are observed (i.e. $\mathscr{\gamma}_{AO}$ demonstrates a slight variation from piston to astigmatism, and compared with small IWA concepts, e.g. phase masks, the APLC demonstrates a roughly similar sensitivity to these aberrations because of its large focal plane mask), so that only the case of tip-tilt ($\mathscr{\gamma}_{AO}$ intermediate case) is presented.

\subsubsection{Offset pointing}
The offset-pointing error refers to the misalignment of the optical axis of the coronagraph with the star. Here, it is assumed that the star is a point source. For instance, with SPHERE the goal is 0.5 mas rms, hence a direct translation of this requirement to an ELT (e.g. 30 to 42 meters telescope), would be a pointing-error residual of about 0.1 mas rms. In practice, the effect of the pointing error is evaluated between 0.1 and 0.5 mas rms to preserve the validity of the analysis to present-day telescopes.

\subsubsection{Pupil shear}
The APLC includes several optical components: apodizer, focal plane mask, and pupil stop.
As a result its performance also depends on the alignment of these components. In particular, the pupil stop has to accurately match the telescope pupil image. This condition is not always satisfied, and the telescope pupil may undergo a significant mismatch that could amount to more than a  few $\%$ of its diameter. It is even more critical for multistage designs because they require several pupil stops in the systems.
The pupil shear is the misalignment of the pupil stop with respect to the telescope pupil image. This is particularly an issue for ELTs, for which mechanical constraints are important for the design. 
Here, a range of pupil shear from 0.1 to 0.5 $\%$ of the pupil diameter is considered. 
Additionally, the optimistic situation where both pupil stops of the MAPLC ($n$=2) are perfectly aligned with each other is assumed, so that only the misalignment of the telescope pupil image is examined.

\subsection{Results}
\label{Results}
Results are presented in Fig. \ref{APLC2error} with the use of azimuthally averaged contrast profiles, where the left column corresponds to a single-stage APLC case ($n$=1), while the right column shows the MAPLC ($n$=2) results.
In all cases, the black curve corresponds to the perfect case (i.e. the reference, no error sources accounted for). It is then straightforward to compare APLC $n=1$ to $n=2$ for different amplitude levels of a given source of error.
This helps to explore the suitability of MAPLC.
\begin{figure}[!ht]
\centering
\includegraphics[width=8cm]{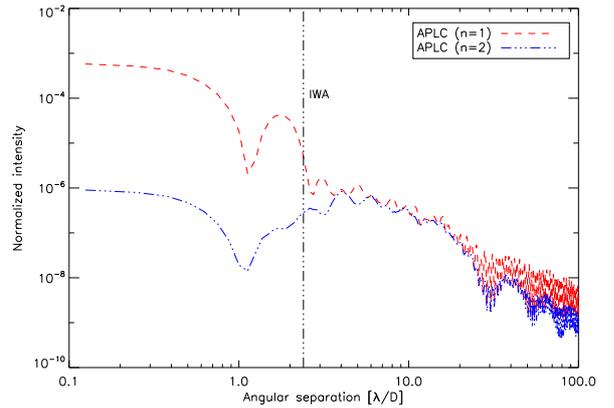}
\caption{Contrast profiles of a single-stage APLC ($n$=1) and a two-stage APLC ($n$=2) obtained in simulation (ELT case) including several error sources as discussed in Sect. \ref{application}. }
\label{ELT}
\end{figure}
\begin{table}[!ht]
\begin{center}
\caption{Preliminary system specification to preserve the advantage of a MAPLC ($n$=2) over an APLC ($n$=1).}
\label{Resum}
\begin{tabular}{ll}
\hline
\hline
Parameter  & Values  \\
\hline
Segment reflectivity ($\%$ ptv)  & $<$1   \\
Segment low-order aberrations\tablefootmark{1} (nm rms)&  $<$50     \\
Offset pointing (mas rms)  & $<$0.3\\
Pupil shear ($\%$)   &  $<$0.2 \\
\hline
\end{tabular}
\tablefoot{The advantage of MAPLC over APLC only concerns angular separations higher than the IWA.
\tablefoottext{1}{Segment aberration levels correspond to XAO-uncorrected values.}}
\end{center}
\end{table} 

In all cases but the offset pointing, APLC two-stages demonstrates a substantial improvement of the energy attenuation in a region confined between the central part of the image (core of the PSF) and the IWA. Although this helps to prevent detector saturation, it does not improve the sensitivity of the instrument for the search of companions.
In all cases, the improvement of APLC $n$=2 over $n$=1 on the coronagraphic halo (i.e. from IWA to further angular separations) is largely uncertain. 
Below, the interest of $n$=2 in the light of each error source is discussed. 

\noindent \textit{Segment reflectivity} -- 1$\%$ (ptv) reflectivity variation already sets a limit on the advantage of $n$=2 configuration from 5$\lambda/D$, while the improvement is limited from IWA to 5$\lambda/D$. From 3$\%$ (ptv) the advantage on the coronagraphic halo of a two-stages APLC has already completely vanished.

\noindent \textit{Segment aberrations} -- Initial segment aberrations of 30 to 50 nm rms (10 to 17 nm rms after correction) already set a similar limitation on the coronagraphic halo to that of a single-stage APLC. Very few improvements are observable for the lower level of these aberrations. For a higher level of segment aberrations, no differences at all are observable between APLC $n$=1 and $n$=2 at higher angular separations than IWA.

\noindent \textit{Offset pointing} -- The impact of the pointing error is dramatic for a MAPLC. A roughly similar level of the coronagraphic halo is delivered compared to that of a single-stage APLC for 0.1 mas rms. Higher values than 0.1 mas rms provide identical contrast levels. 
While for most error sources the improvement remains important and stable before IWA, the energy reduction at shorter angular separations than the IWA degrades even more as the offset gets higher. 

\noindent \textit{Pupil shear} -- From 0.2$\%$ pupil shear, contrast profiles are comparable. As the internal misalignment between the two pupil-stops of the two-stages APLC is neglected (the simulation assumes identical misalignment of both pupil stops to the telescope pupil image), in practice the pupil shear will likely further degrade the performances. 

\subsection{Combining error sources}
\label{application}
In this monochromatic simulation all parameters discussed above are combined, assuming reasonable values: an initial (prior to XAO correction) of 36 nm rms piston and tip-tilt, 50 nm rms defocus and astigmatism, 5$\%$ (ptv) segment reflectivity, 0.1 mas rms offset pointing, and 0.2$\%$ pupil shear. \\
It is found that APLC $n$=2 and $n$=1 provide undistinguished contrast profiles from 4$\lambda/D$ (Fig. \ref{ELT}). 
Although further energy attenuation is observable with a two stages APLC from the core of the image to the IWA (Fig. \ref{ELT}), no improvement occurs in the halo (basically at higher angular separations than the IWA).   

\section{Laboratory experiment}
The previous section has provided parametric sensitivity analysis results through a numerical treatment, and a preliminary insight of a system level specification to preserve the interest of a MAPLC ($n$=2) is shown in Table \ref{Resum}.
To show the relevance of these results to a real situation, the results obtained in the lab with an APLC one- and two-stages are discussed below.  

\subsection{Experimental conditions}
The optical setup is a near-infrared coronagraphic transmission bench developed at ESO in 2008, and lates dismounted to feed the near-IR optical path of HOT, the High-Order Testbench (adaptive optics facility at ESO).
The main characteristics of the optics of the first stage (from the light source to the focal plane of the second Lyot-mask) are described in several papers \citep[e.g.][]{microdots, microdots2}. The second stage (from the second Lyot mask plane to the detector plane) makes use of similar optical components as the first one (e.g. same lenses, $\textit{f}$-number...). Focal plane masks were installed at an F/48.4 beam. Re-imaging optics were made with IR achromatic doublets. The IR camera used (the Infrared Test Camera) uses a HAWAII $1k\times1k$ detector, cooled to $105^{\circ}$ K with a vacuum pressure of $10^{-5}$ mbar. Internal optics were designed to reach a pixel scale of 5.3 mas ($\sim$8 pixels per $\lambda/D$). The experiment was performed in $H$-band using a narrowband filter ($\Delta \lambda/\lambda =1.4\%$). The use of the narrow-band filter ensures that chromaticity does not impact the performance of MAPLC compared to other error sources, as in the simulations presented here. 

The optical setup was designed to simulate the 8 m VLT pupil. The 3mm entrance aperture diameter ($\Phi$) was made in a laser-cut, stainless-steel sheet to an accuracy of 0.002 mm.
The central obscuration was scaled to be 0.47 mm $\pm$ 0.002 mm and the spider-vane thickness is 60$\mu$m $\pm$ 4$\mu$m (15$\mu$m is indeed the nominal size, but not available during the experiment). 
The APLC is a 4.5$\lambda/D$ diameter (small mask size regime, i.e. $\lesssim$ 5$\lambda$/D).
The pupil-stops mimic the VLT pupil mask with equivalent spider-vane thickness (60$\mu$m $\pm$ 4$\mu$m), an outer diameter that is reduced by a factor 0.96$\times \Phi$ (2.88 mm $\pm$ 0.002 mm), and a central obscuration equals to 0.16$\times \Phi$ (0.49 mm $\pm$ 0.002 mm). 
\begin{figure}[!ht]
\centering
\includegraphics[width=8cm]{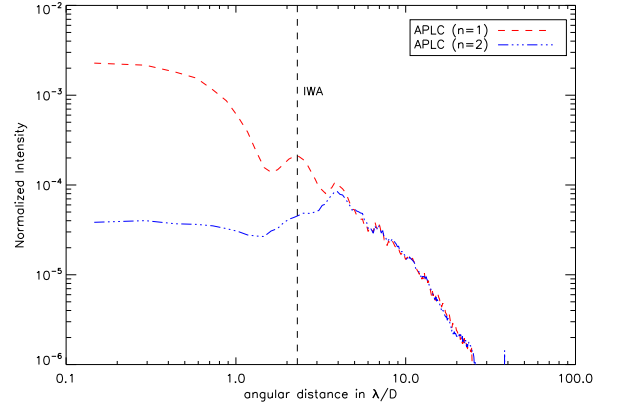}
\caption{Contrast profiles of a single-stage APLC ($n$=1) and a MAPLC ($n$=2) obtained in the lab in $H$-band with a VLT-like pupil.}
\label{labo}
\end{figure}
\begin{figure}[!ht]
\centering
\includegraphics[width=8cm]{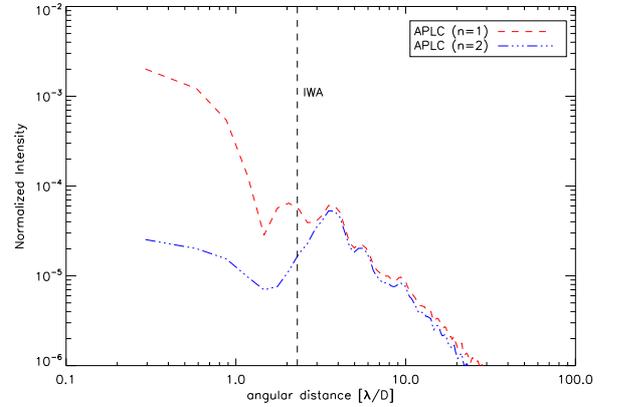}
\caption{Simulated  contrast profiles of a single-stage APLC ($n$=1) and a MAPLC ($n$=2) assuming the experimental conditions.}
\label{labo2}
\end{figure}
The pupil-stop throughput is about 90$\%$, and the Strehl ratio of the bench is of about $\sim$92$\%$.
\begin{figure*}[!ht]
\centering
\includegraphics[width=4.5cm]{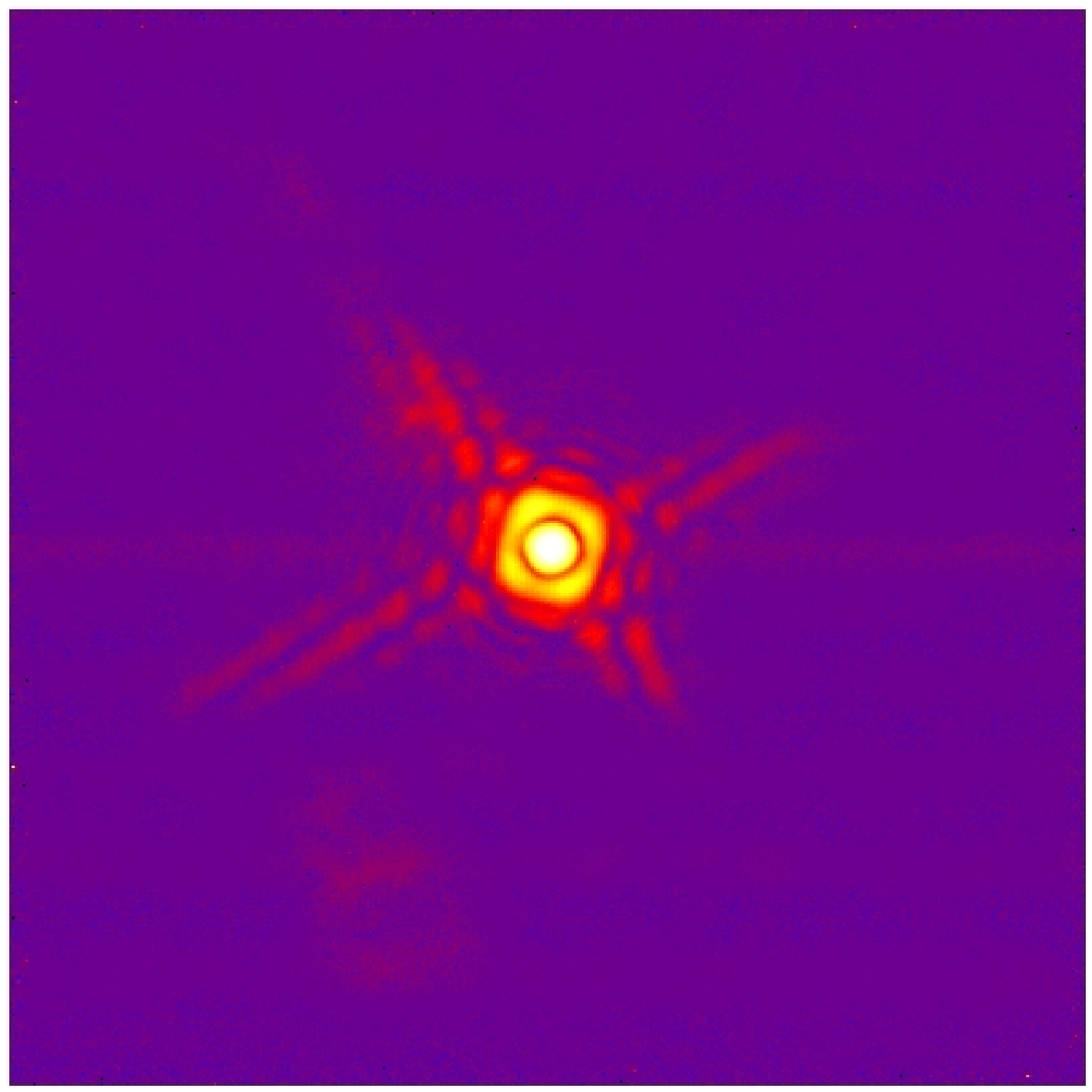}
\includegraphics[width=4.5cm]{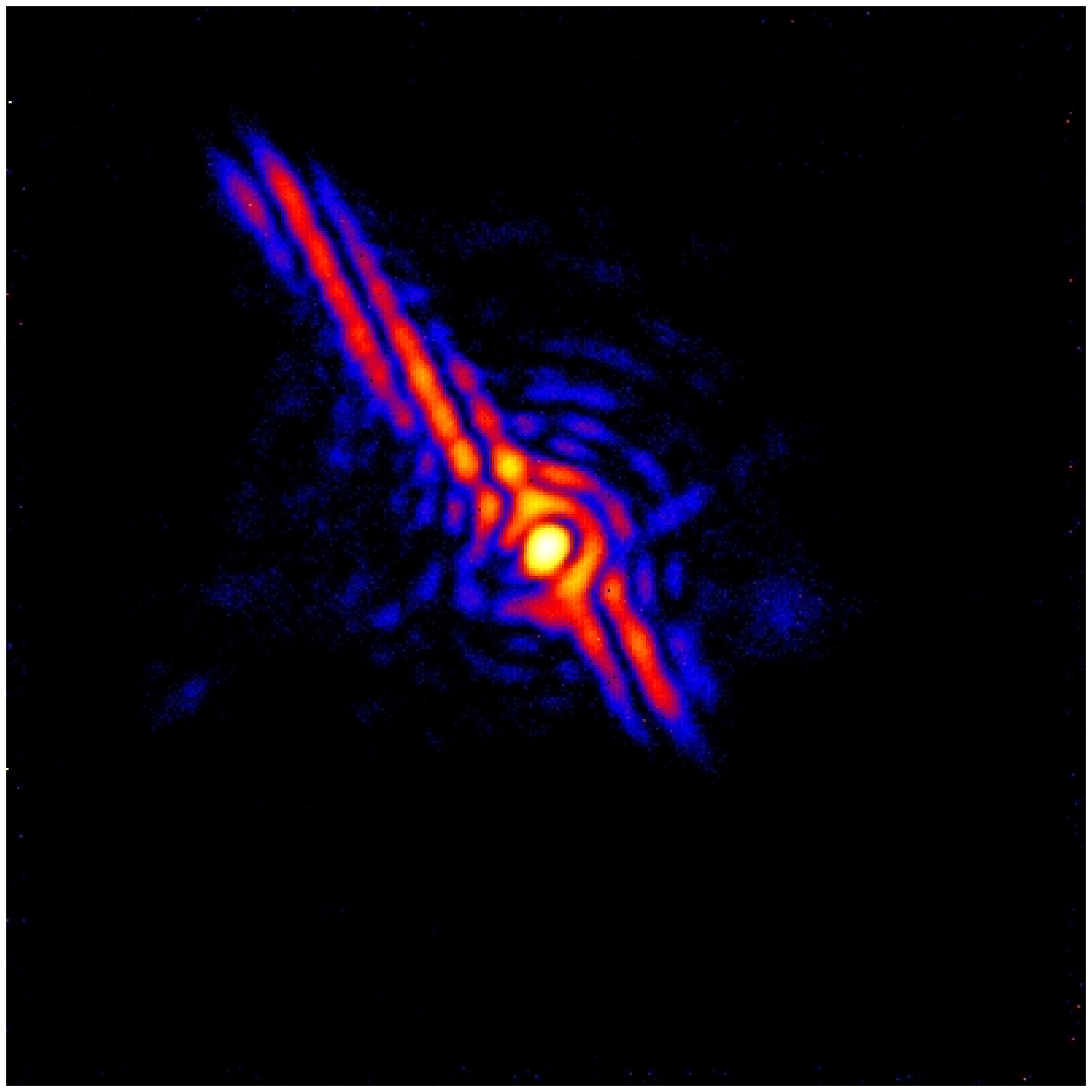}
\includegraphics[width=4.5cm]{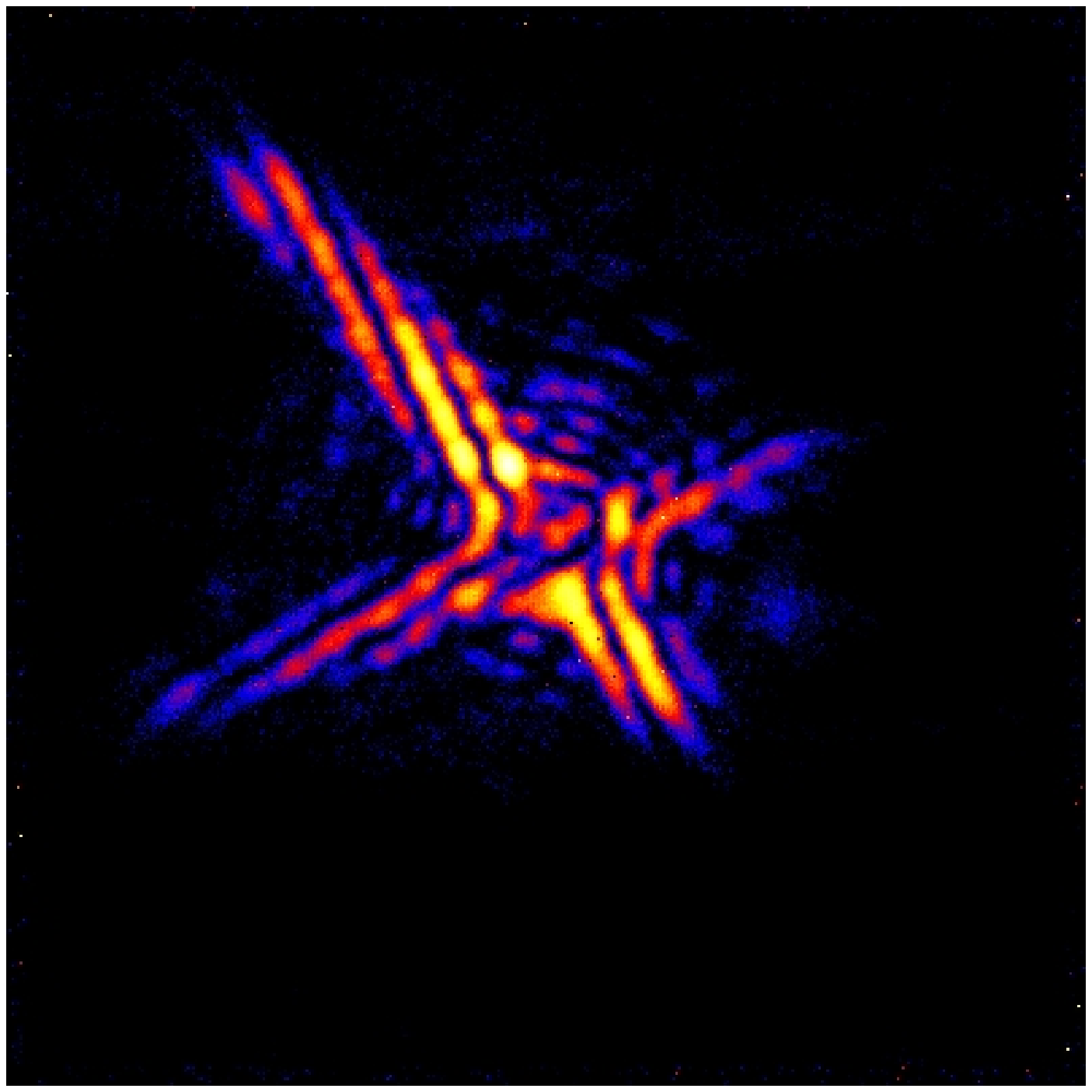}
\caption{$H$-band images recorded during the experiment, from the left to the right: apodized PSF, APLC ($n$=1) and APLC ($n$=2). Colors and contrast are arbitrary for the sake of clarity.}
\label{laboImage}
\end{figure*}

\subsection{Results}
Monochromatic contrast profiles are presented in Fig. \ref{labo}, while images recorded during the experiment are shown in Fig. \ref{laboImage}.
Because the spider vanes thickness of the telescope pupil and the pupil stops are identical, and considering the small size of the pupil planes ($\Phi=$3mm), a fine alignment was difficult to reach (as exhibited in Fig. \ref{laboImage}, where diffraction residuals from the spider vanes are not homogeneously distributed across the coronagraphic images).  
Contrast profiles presented in Fig. \ref{labo} are azimuthally averaged so that the impact of the diffraction residuals from the spider vanes is diminished. 
Additionally, a contrast evaluation in an area free of these residuals does not provide different results. 

Although misalignment errors (pupil shear) play a role, the residual speckles appear to be the major limitation on the coronagraphic halo. 
Undistinguished contrast profiles are found between a single-stage APLC and a two-stages APLC. Improvement is only observable from the core to the IWA, as expected from the simulation provided in the previous section.
Results obtained in the experiment (VLT-pupil) and in the simulation presented in Sect. \ref{application} in the context of an ELT including several error sources are very similar, which further confirms the numerical analysis (i.e. the entrance pupil configuration does not impact MAPLC performances/behavior, while IWA are similar). 

Additionally a simulation has been carried out assuming experimental conditions (Fig. \ref{labo2}), and supports the results obtained in the laboratory (Fig. \ref{labo}).
The simulation assumes similar spectral bandwidth, ideal coronagraphic components (i.e. not as-manufactured), 0.2$\%$ pupil shear, $0.5^{\circ}$ pupil rotation (spider vanes presence on the pupil), and wavefront errors of the optical components prior to the last pupil stop (wavefront high-frequency components, estimated to 25 nm rms) using a typical $f^{-2}$ power law of manufactured optics for the PSD. 
Although the contrast levels are slightly different between simulation and experiment, the behavior of the MAPLCs is found to be identical. \\
The small discrepancy in contrast may be caused by several limitations of the simulation. 
The simulation does not take into account the as-manufactured coronagraphic components, the likely misalignment of the apodizer with the pupil, and FPM with the star (x-, y- and focus), the misalignment of the two pupil-stops with each other, and assumes optimistic levels of pupil shear and pupil rotation. It is also neglected that the pupil stops are not black-coated (e.g. anodization), which may produce potential reflections. Additionally, the optical setup is not covered by panels, and thus cannot be shielded from the room turbulence. Even if it is difficult to ascertain the origin of the small discrepancy,  the simulation presented here provides a qualitative support, while quantitative aspects must be carefully considered.

\section{Conclusion}
With the future worldwide rise of extremely large telescopes, the detection of rocky planets (e.g. super-earths) comes as a major scientific goal, which implies the study and development of advanced starlight-cancellation techniques.
To reach the required high contrasts, improvement of a coronagraph efficiency is mandatory. 
Although, the observation and data reduction strategies (e.g. multi-wavelength imaging speckle rejection techniques) can relax the constraints on the efficiency of a coronagraph at the level of the residual atmospheric halo, 
better contrast capabilities would be preferably not limited by pinned speckles \citep{SoummerSpeckles07}.
With these considerations in mind, the study has been carried out at the contrast level where the limitation comes from the coronagraph. 

Multistage APLC were found inappropriate for ELTs. Although a large improvement on the peak -- at least from the core of the PSF to the IWA -- is achievable, avoiding the detector saturation (longer integration time), no improvement on the halo is observed when considering realistic error sources. As a result, MAPLC does not increase the instrument sensitivity to detect a fainter companion. While we have selected a few error sources, the study could be extended to others (e.g. finite size of the star, static aberrations upstream and downstream of the coronagraph, apodizer and FPM misalignments...), but they are not expected to alter its outcome.
The results presented here were principally aimed for ELTs but are indeed, by the nature of some parameters investigated (e.g. offset pointing, pupil shear), directly applicable to present-day telescopes (non-segmented telescopes, e.g. 8-10 m telescope). Additionally, the simulations are further confirmed by a laboratory experiment in the near-IR enabling the comparison of a single and double-stages APLC in a real environment . 
The APLC remains a promising concept for ELTs \citep{Corono}, but in its conventional single-stage implementation. 

Most of the coronagraphs can potentially be implemented in the form of multistage designs. The conclusion presented here can therefore be extended to any of these concepts, unless either achromatization \citep[e.g. multi-FQPMs,][]{2007lyot.confE..25B}, or significant IWA reduction for more audacious science cases are the ultimate goals. In these regimes, post-coronagraphic wavefront error residuals must be tackled to reach high-contrast level requirements for future ELT-planet-hunter instruments. 

\acknowledgements
This research has been partially funded as part of the European Commission, Sixth Framework Programme (FP6), ELT Design Study, Contract No. 011863, Seventh Framework Programme (FP7), Capacities Specific Programme, Research Infrastructures; specifically the FP7, Preparing for the construction of the European Extremely Large Telescope Grant Agreement, Contract number INFRA-2007-2.2.1.28.
P. M thanks A. Boccaletti for his comments and suggestions on the paper.
\bibliography{MyBiblio}

\end{document}